# Community structures in simplicial complexes: an application to wildlife corridor designing in Central India - Eastern Ghats landscape complex, India

**Saurabh Shanu[1], Shashankaditya Upadhyay[2], Arijit Roy[3], Raghunandan Chundawat[4], Sudeepto Bhattacharya[5*]**

[1]Department of Virtualization, School of Computer Science and Engineering, University of Petroleum and Energy Studies, Dehradun 248007, Uttarakhand, India

[2]Department of Electrical Engineering, Indian Institute of Technology, Delhi, New Delhi, India 110016

[3]Disaster Management Science Department, Indian Institute of Remote Sensing, Indian Space Research Organization (ISRO).4, Kalidas Road, Dehradun 248001, Uttarakhand, India

[4]BAAVAN, S-17 Panchsheel Apartment, A-1 Block Panchsheel Enclave, New Delhi 110 017, India

[5]Department of Mathematics, School of Natural Sciences, Shiv Nadar University, P.O. Shiv Nadar University, Greater Noida, Gautam Buddha Nagar 201314, Uttar Pradesh, India

[*]Author for correspondence. E-mail: sudeepto.bhattacharya@snu.edu.in

**Abstract**

The concept of simplicial complex from Algebraic Topology is applied to understand and model the flow of genetic information, processes and organisms between the areas of unimpaired habitats to design a network of wildlife corridors for Tigers (Panthera Tigris Tigris) in Central India Eastern Ghats landscape complex. The work extends and improves on a previous work that has made use of the concept of minimum spanning tree obtained from the weighted graph in the focal landscape, which suggested a viable corridor network for the tiger population of the Protected Areas (PAs) in the landscape complex. Centralities of the network identify the habitat patches and the critical parameters that are central to the process of tiger movement across the network. We extend the concept of vertex centrality to that of the simplicial centrality yielding inter-vertices adjacency and connection. As a result, the ecological information propagates expeditiously and even on a local scale in these networks representing a well-integrated and self-explanatory model as a community structure. A simplicial complex network based on the network centralities calculated in the landscape matrix presents a tiger corridor network in the landscape complex that is proposed to correspond better to reality than the previously proposed model. Because of the aforementioned functional and structural properties of the network, the work proposes an ecological network of corridors for the most tenable usage by the tiger populations both in the PAs and outside the PAs in the focal landscape.

**Keywords:** Centrality measures, Community detection, Corridor, Ecological networks, Landscape complex, Simplicial complex,

# 1. Introduction

Wildlife Corridors within a landscape manifests the integration of natural and anthropogenic disturbances with the features in the landscape and its landforms that facilitate landscape linkages [25]. Thus, the landscape linkages represented by corridors, formally explain the extent to which the landscape obstructs or eases the movement of species between the territorial regions occupied by the species. [37, 38]. Hence, we state that corridors are stretches of successive habitats, usually linear, embedded within a landscape matrix, which joins two or more habitat patches. Corridors provide a landscape link between the habitats and proposed for conservation, implies that it would enhance or maintain the feasibility of certain wildlife populations in the territorial regions. Further, we understand the passage or movement as travel via a corridor by individuals of focal species from an occupied territory to another territory. Wildlife corridors, as inferred from the discussions, are key elements of a landscape, which connect fragmented regions of the landscape. [5, 11, 10].

In this work, we focus on the animal movements across the landscape linkages and henceforth term these linkages as wildlife corridors. Animal species differ greatly in terms of habitat extent with underlying specialist as well as generalist adaptation and their adaptability to habitat disruption and change [13]. In this scenario, the specific role of landscape level connectivity as a conservation strategy is to integrate habitats into a network system. This would propose and restore the natural flow of information (genes) and energy (food chain) between distributed populations of species, separated due to various land use patterns, by supporting the spatial-temporal movements of both living and non-living processes within the landscape complex [1, 4, 5, 10, 13, 20, 23, 25, 32, 35].

The recent ongoing studies suggest that wildlife corridor designs must be species centric and consider the impacts of migration. We strongly argue and intend to establish that the models for corridor design must be forming complex network of corridors and develop community structure amongst the territorial regions. A community structure is obtained within a network when the vertices of the network potentially overlap or can be easily collected into a set of vertices portraying a dense interconnection of elements in each set of vertices [51].

Various models and strategies have emerged in the last few decades for designing wildlife corridors. The most prominent and accepted strategies used to design wildlife corridors follow either the circuit theory [30] or the concepts of Minimum Spanning Tree [34]. Both the concepts justify and convince for, "What could be the best path which would support the movement of wild animals in a given landscape." They also suggest that a detailed topographic dataset and habitat analysis of animals assist in designing the migration corridors. However, they do not influence the animals to move only in the designed corridors. Hence, it is not necessary that animals would move only in the designed corridors. Subject to availability and satisfaction of ecological necessities, animals may even traverse the landscape along routes that essentially do not capture the features of most optimal corridors according to the designing principles of the applied theories [4]. Field studies confirm that migrating animals move out of the corridors suggesting that the two proposed solutions discussed above for corridor design do not capture all the available corridors for the animal movement. In addition, the proposed solutions are not able to answer or justify about the behavioural pattern of species in and around the corridors, which takes into account the resident territorial populations in the landscape as well as the tiger population in the PAs.

In a recent work, the authors following the minimal spanning tree approach to design wildlife corridor in a given landscape have used the Kruskal's [34], where for $n$ vertices in the landscape there existed $n-1$ corridors. The analysis was applied over a graph $G = (V, E)$, where $V$ represents the set of vertices and $E$ represents the set of edges. For wildlife corridor design, the Protected Areas (PAs) serve as the elements of the set of vertices $V$ and the corridors between them as the set of edges $E$. Using pairwise interaction between the habitats (the PAs), the work provided an optimal solution to the the central problem of designing a viable wildlife corridor. The pairwise interaction was able to find only one path, the most optimal path between any two vertices and did not support multiple connections largely. As a logical next step to the cited work, we would like to model tiger corridors that would correspond even better to reality. Towards this objective, in this work we propose a method for designing a corridor network for tigers that reside within the PAs in the landscape as well as those individuals and populations that reside in the territorial forests in proximity with the PAs. The later mainly comprise individual animals [40]. Multi-body interactions instead of the pairwise interactions capture the interactions between the tiger habitats inside as well as outside the PAs where the tigers may traverse in the focal landscape. To model these interactions, we use simplicial complex, a mathematical object from the field of algebraic geometry. We initially identify the communities to represent sets of PAs, which follow congruent features. Understandings of species behavior and the species transition patterns obtained by identifying and studying such communities could prove to be instrumental in further deepening our understanding of species. Finally, informed about the possible communities and geographical constraints, we provide a simplicial network of the community structure, which we claim, could serve as a model corridor network. In the present work, we present a higher-level computational approach for designing corridor for the tigers in the Central India-Eastern Ghats landscape complex

Section 2 describes in brief the key mathematical concepts used for the work, followed by Sections 3 that brings out the results of work conducted in the Minimum Spanning Tree concepts for tiger corridors, 4 and 5 describing the modelling and the conclusion of the work, respectively.

## 2. Centrality measures and Simplicial Complex

To make the paper self-contained, this section begins with a brief introduction to the foundational definitions required for developing the proposed theory further.

Networks offer a formal and comprehensive description of complex systems. Usually applied to model experimental information where a multi-body interaction is of importance and evolves over certain domain. For the problem in this paper, the relationship between the tiger and the functional parameters of a landscape that either support or obstruct the tiger movement, have importance and may evolve with space and time. Hence, to model such complex systems we adopt a perspective using the theory of complex networks for modelling.

A network $N$ is a four tuple $(V_\lambda, E_\lambda, \psi_\lambda, \Lambda)$ with an algorithm $\beta$ such that for $\Lambda \neq \phi, k \in \Lambda, V_\lambda$ is a set of vertices $V_k$, $E_\lambda$ is a set of edges $E_k$, $\psi_\lambda$ is incidence function $\psi_k : E \to |V|^2$ where $|V|^2$ is the set of not necessarily distinct unordered pairs of vertices such that $(V_k, E_k, \psi_k)$ is a graph given by the algorithm $\beta(k)$. The incidence function $\psi$ provides structure to a graph by associating to each edge an unordered pair of vertices in the graph as $\psi(x) = \{v_i, v_j\}: v_i, v_j \in V, \forall\, x \in E \subseteq [V]^2$. Here $k$ is the temporal component by virtue of which a network can evolve as per the given algorithm $\beta$ [39].

A graph is an algebraic object such that an unlabelled graph represents an isomorphism class of otherwise labelled graphs. We call a network as static network if the temporal component $\Lambda$ consist of a single element $i$, otherwise the network is a dynamic network. For the purpose of our work in this paper, we define an ecological network as a network $N$ in which $V_\lambda$ is the set of habitat patches for the tigers, and $E_\lambda$ is the set of paths of ecological matter between two distinct habitat patches [39]. From the above discussions, we can deduce that a graph consists of vertices connected by edges, while a network consists of nodes connected by links, where the links and nodes depend on the incidence function and the spatial-temporal evolutions [52]. In this paper, we shall use the terms graph and networks interchangeably.

Centrality measures, originally developed as a basic framework for using network and graph theory for exploring social structures [2, 3], play a vital role in structural analysis. Centrality measures are structural measures and have recently gained extensive importance in the analysis and designing of ecological networks [9, 10, 16]. The inbuilt concept of a centrality measure is that it ranks the vertices $V$ or the edges $E$ of a graph $G$ by assigning real values based on the importance of vertices [8]. Thus, the centrality measures help us to obtain the central components of a graph. A center of $G$ is a vertex of $G$ such that maximum degree of the vertex is as small as possible [7]. However, the centrality measures are influenced by the structure of the graph. The following definition of a structural index states the underlying application:

**Structural index.** Let $\Gamma_1\big(V(\Gamma_1), E(\Gamma_1), \Psi_{\Gamma_1}\big)$ and $\Gamma_2\big(V(\Gamma_2), E(\Gamma_2), \Psi_{\Gamma_2}\big)$ be two graphs and let $X$ represent the set of vertices or edges of $\Gamma_1$, $G$ and $H$ represent two sub graphs of $\Gamma_1$. Then, $s: X \to \mathbb{R}$ is called a structural index if and only if the following condition is satisfied: $\forall x \in X: G \cong H \Longrightarrow s_{\Gamma_1}(x) = s_{\Gamma_2}(\phi(x))$, where $\phi: V(\Gamma_1) \to V(\Gamma_2)$ is an isomorphism, and $s_{\Gamma_1}(x)$ denotes the value of $s(x)$ in $\Gamma_1$ and $s(\phi(x))$ denotes the value of $s(x)$ in $\Gamma_2$ [34, 39, 53].

Nominally, a centrality measure $c$ induces at least a semi-order on the set of vertices or edges of the graph in consideration as is required to be a structural index. Thereby, we say $x \in X$ is at least as central as $y \in X$ if $c(x) \geq c(y)$ [39].

Various centralities are used for the purpose of real world problems and respective modellings. In the succeeding paragraphs, we discuss about the centralities that we have used for this paper.

We use the degree centrality ($DC$) to obtain the significance of a vertex in a graph based on its degree, eigenvector centrality ($EC$) to rank the vertices in order of relevance of usefulness, betweenness centrality ($BC$) to obtain the functional minimal cost paths between a the pair of vertices, closeness centrality ($CC$) to obtain essential vertices of the graph as being measured by how close it is structurally to all other vertices in the graph, and for analysis of the tiger corridor network we use subgraph centrality ($SC$), positively scaled subgraph centrality ($SP_i$), and negatively rescaled subgraph centrality ($SN_i$). Additionally, based on edge-btweenness centrality we used the Newman – Girvan algorithm for detecting communities in the network. Independent communities act as if meta-vertices in the network, which help us to obtain a wide-ranging map of a network, making its study simpler. They often have very different features than the general features of the networks. Thus, only working on the general features usually leaves many significant and interesting features inside the networks [29]. In addition, being able to identify these communities within a network, insights into how network function and network structure affect each are calculated and obtained [42].

**Newman – Girvan algorithm.** Within a network, for the purpose of detecting and identifying communities, we use the Newman – Girvan algorithm. [28]. Edge-betweeness centrality is the basis on which the algorithm works. Edge betweeness centrality is analogous to the standard betweeness centrality applied only to edges. Formally, for an edge $e$ and the pair of vertices $(s, t)$ the betweeness centrality is given by

$$BC(e) = \sum_{s \neq v \neq t} \frac{\sigma_{st}(e)}{\sigma_{st}}$$

Where $\sigma_{st}$ denotes the number of functional shortest paths between $s$ and $t$, and $\sigma_{st}(v)$ denotes the number of such paths passing through $e$. The idea behind the algorithm is that if a network contains communities or groups that only loosely connected by a few intergroup edges, then all shortest paths between different communities must go along one of those few edges, and such edges will have high edge-betweenness. By removing these edges, the groups stay separated from one another to reveal the underlying community structure of the network [26, 27, 28, 29, 34].

**Simplicial Complex.** It is a quotient space of a collection of disjoint simplices obtained by identifying certain of their faces via the canonical linear homeomorphisms, which preserve the ordering of vertices [14, 19]. A simplicial complex $S$ may be defined as a set of simplices such that if a simplex $P$ is an element of the set $S$ then all faces of $P$ are also elements of $S$. To capture the essence of simplicial complexes in Complex networks, a defined dimensional space happens to be of key importance. Thus a $k$-simplex is a mathematical object with $(k + 1)$ vertices which exists in a $k$-dimensional space [14, 19]. A set of simplices constitutes the Simplicial Complex For example if $A = \{a_0, a_1, a_2, \ldots, a_k\}$ creates a simplex then all its faces $F = \{a_0, \ldots, a_{i-1}, a_{i+1} \ldots, a_k\}$ also create a simplex. Further all the faces of $F$, $F` = \{a_0, \ldots, a_{i-1}, a_{i+1}, \ldots, a_{j-1}, a_{j+1}, \ldots a_k\}$ also create simplex until we reach 0-simplices formed just by the nodes [14, 19].

**Clique Complex.** A clique complex can be obtained from a network. The set of the network become the set of the simplicial complex. Let $Z$ be a clique of $n$ vertices in the network. Then, $Z$ is a $(n - 1)$-simplex in the clique complex. As an example in Fig 1 we describe a simplicial complex which has one 3-simplex {$a_0$, $a_1$, $a_2$, $a_3$}, and six 2-simplices {$a_0$, $a_1$, $a_2$}, {$a_0$, $a_1$, $a_3$}, {$a_0$, $a_2$, $a_3$}, {$a_1$, $a_2$, $a_3$}, {$a_2$, $a_3$, $a_4$}, and {$a_3$, $a_4$, $a_5$}. It also has eleven 1-simplices represented by the edges and seven 0-simplices, the vertices [14, 19].

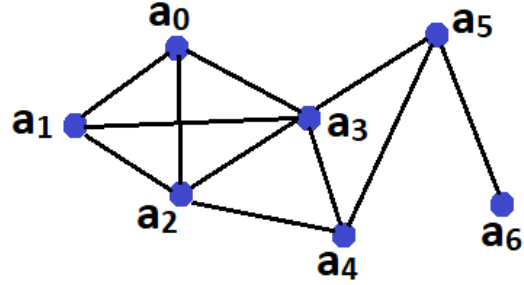

**Fig 1. A simplicial complex with labeled vertices**

In network conjecture, it is clear when two vertices are adjacent. However, adjacency is a tricky to define in simplicial complexes. The adjacency of two *n*-simplices *p* and *q* can be defined in two ways: lower and upper adjacency.

***Definition 1***. Let *p* and *q* be two *n*-simplices. Lower adjacency exists between the two *n*-simplices if they share a common face. Which implies, for two distinct *n*-simplices $p = \{p_0, p_1... p_k\}$ and $q = \{q_0, q_1... q_k\}$, *p* and *q* are lower adjacent if and only if there is a $(n-1)$-simplex $\beta = \{r_0, r_1... r_{k-1}\}$ such that $\beta \subset p$ and $\beta \subset q$. We denote lower adjacency by $p \smile q$ [14]. In the simplicial complex in Fig 1, the 2-simplices $\{a_0, a_1, a_3\}$ and $\{a_1, a_2, a_3\}$ are lower adjacent because they share a common 1-simplex $\{a_1, a_3\}$ which is a common face for both. So we can write $\{a_0, a_1, a_3\} \smile \{a_1, a_2, a_3\}$[14, 19].

***Definition 2***. Let *p* and *q* be two *n*-simplices. Then the two *n*-simplices are upper adjacent if they are both faces of the same common $(n+1)$-simplex. That is, for $p = \{p_0, p_1... p_k\}$ and $q = \{q_0, q_1... q_k\}$ then *p* and *q* are upper adjacent if and only if there is a $(n+1)$-simplex $\lambda = \{r_0, r_1... r_{k+1}\}$ such that $p \subset \lambda$ and $q \subset \lambda$. We denote the upper adjacency by $p \frown q$ [14]. In the simplicial complex in Fig 1, the 1-simplices $\{a_2, a_4\}$ and $\{a_3, a_4\}$ are upper adjacent because they are both faces of the 2-simplex $\{a_2, a_3, a_4\}$ which is a common face for both. So we can write $\{a_2, a_3\} \frown \{a_3, a_4\}$[14, 19].

This work is a succession of the work in [34] to generate a solution for tiger corridor networks that is both optimal and complete. Hence, for the sake of providing the basis for our arguments and deductions we mention the following theorem.

***Theorem 1***. *(The Completeness Theorem). If T is a theory and every model of T satisfies* φ *, then* $T \models \varphi$ *where* φ *represents a set of formulas* [55].

## 3. Preamble to the present problem

In order to ensure logical connectivity and flow of arguments, in this section we reproduce certain relevant results and gist of some discussions from [34]. .

**Table 1: Protected Areas in Central India Eastern Ghats Landscape [34]**

| S.No | Tiger habitat | Code |
|---:|---|---|
| 1. | Sariska | 1 |
| 2. | Ranthambore | 2 |
| 3. | Kuno-Shivpur- Madhav | 3 |
| 4. | Raisen | 4 |
| 5. | Indore-Dewas | 5 |
| 6. | Satpura | 6 |
| 7. | Melghat | 7 |
| 8. | Bor | 8 |
| 9. | Tadoba | 9 |
| 10 | Shayadri | 10 |
| 11 | Srisailam | 11 |
| 12 | Adilabad | 12 |
| 13 | Nagzira | 13 |
| 14 | Baranwapara | 14 |
| 15 | Satkosia | 15 |
| 16 | Simlipal | 16 |
| 17 | Achanakmar | 17 |
| 18 | Palamou | 18 |
| 19 | Sanjay-Dubri-Guru Ghasidas | 19 |
| 20 | Bhandavgrah | 20 |
| 21 | Kanha | 21 |
| 22 | Panna | 22 |
| 23 | Pench | 23 |

The PAs shown in Table 1 were vertices for the graph in the focal landscape and on application of 2-persons Prisoner's Dilemma game [53], based on the payoffs, between tiger and the parameters of the landscape; a weighted graph was obtained [34]. On implementation of Kruskal's algorithm, we obtain the Minimum Spanning Tree for the tiger corridor network in focal landscape as shown in Fig 2.

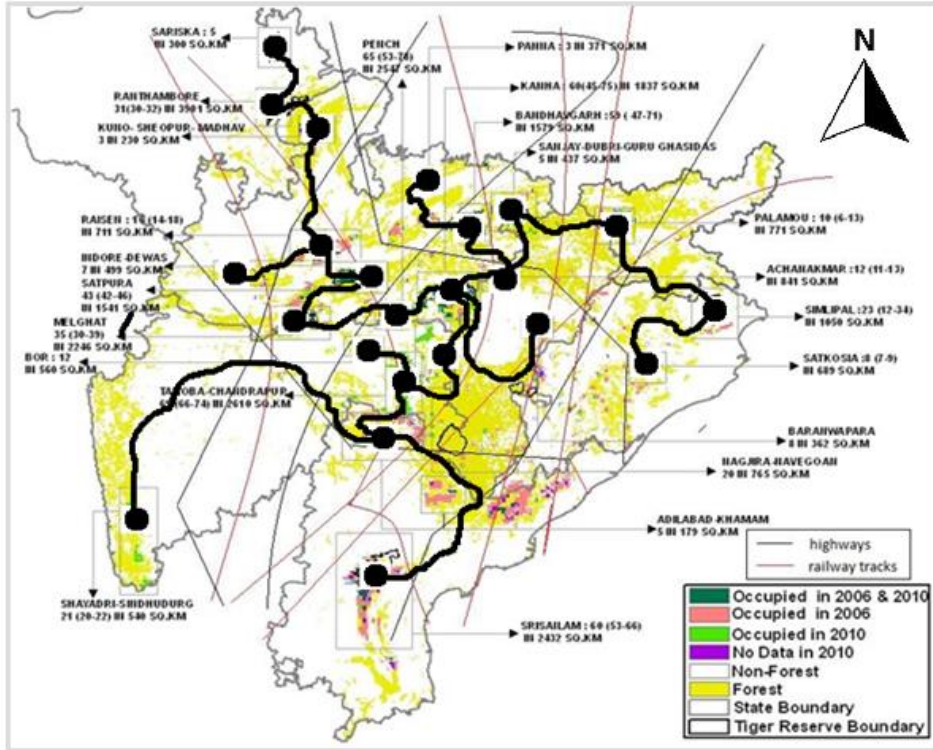

**Fig 2. The feasible tiger corridor network, given by a MST using Kruskal's algorithm, overlaid on the map of the focal landscape complex [34].**

On centrality analysis, potentially important patches or vertices were obtained. A table of vertices (habitat patches) ranked (highest-to-lowest) by various centrality measures is given below:

| DC | EC | BC | CC | SC | SP$_1$ | SP$_2$ | SP$_3$ | SP$_4$ | SP$_5$ | SP$_6$ | SN$_1$ | SN$_2$ | SN$_3$ | SN$_4$ | SN$_5$ | SN$_6$ |
|---|---|---|---|---|---|---|---|---|---|---|---|---|---|---|---|---|
| 6 | 23 | 21 | 21 | 23 | 23 | 21 | 21 | 21 | 21 | 21 | 23 | 23 | 23 | 23 | 23 | 23 |
| 21 | 21 | 3 | 6 | 21 | 21 | 6 | 6 | 6 | 6 | 6 | 21 | 21 | 21 | 21 | 21 | 21 |
| 23 | 6 | 14 | 13 | 6 | 6 | 20 | 20 | 20 | 20 | 20 | 6 | 6 | 6 | 6 | 6 | 6 |
| 7 | 13 | 13 | 23 | 9 | 9 | 22 | 22 | 22 | 22 | 22 | 9 | 9 | 13 | 13 | 13 | 13 |
| 9 | 9 | 22 | 14 | 13 | 13 | 23 | 17 | 17 | 17 | 17 | 13 | 13 | 9 | 9 | 9 | 9 |
| 13 | 7 | 6 | 22 | 7 | 7 | 14 | 14 | 4 | 4 | 4 | 7 | 7 | 7 | 7 | 7 | 7 |
| 14 | 8 | 2 | 20 | 17 | 17 | 13 | 7 | 7 | 7 | 7 | 8 | 8 | 8 | 8 | 8 | 8 |
| 17 | 14 | 12 | 17 | 8 | 14 | 7 | 13 | 13 | 18 | 18 | 17 | 17 | 17 | 14 | 14 | 14 |
| 4 | 10 | 20 | 7 | 14 | 8 | 17 | 4 | 18 | 13 | 13 | 14 | 14 | 14 | 17 | 17 | 10 |
| 8 | 17 | 4 | 5 | 10 | 10 | 18 | 23 | 14 | 14 | 14 | 10 | 10 | 10 | 10 | 10 | 17 |
| 10 | 12 | 7 | 9 | 12 | 12 | 4 | 18 | 23 | 23 | 10 | 12 | 12 | 12 | 12 | 12 | 12 |
| 12 | 20 | 23 | 4 | 20 | 18 | 9 | 9 | 10 | 10 | 9 | 20 | 20 | 20 | 20 | 20 | 20 |
| 18 | 5 | 17 | 12 | 4 | 4 | 12 | 10 | 9 | 9 | 23 | 4 | 4 | 22 | 22 | 5 | 5 |
| 20 | 22 | 9 | 8 | 18 | 20 | 10 | 12 | 15 | 16 | 19 | 18 | 22 | 4 | 5 | 22 | 22 |
| 22 | 4 | 5 | 19 | 22 | 22 | 8 | 15 | 12 | 15 | 16 | 22 | 18 | 5 | 4 | 4 | 4 |
| 3 | 18 | 15 | 3 | 5 | 5 | 3 | 16 | 16 | 19 | 15 | 5 | 5 | 18 | 18 | 18 | 18 |
| 5 | 19 | 10 | 10 | 19 | 19 | 5 | 19 | 19 | 12 | 12 | 19 | 19 | 19 | 19 | 19 | 19 |
| 19 | 3 | 18 | 18 | 3 | 3 | 19 | 3 | 8 | 8 | 8 | 3 | 3 | 3 | 3 | 3 | 3 |
| 2 | 15 | 8 | 15 | 15 | 15 | 15 | 8 | 3 | 3 | 3 | 15 | 15 | 15 | 15 | 15 | 15 |
| 15 | 11 | 1 | 2 | 16 | 16 | 16 | 5 | 5 | 5 | 5 | 16 | 16 | 16 | 16 | 16 | 16 |
| 16 | 16 | 11 | 11 | 2 | 2 | 2 | 2 | 2 | 2 | 11 | 2 | 11 | 11 | 11 | 11 | 11 |
| 1 | 2 | 16 | 16 | 11 | 11 | 11 | 11 | 11 | 2 | 11 | 2 | 2 | 2 | 2 | 2 |
| 11 | 1 | 19 | 1 | 1 | 1 | 1 | 1 | 1 | 1 | 1 | 1 | 1 | 1 | 1 | 1 | 1 |

**Fig. 3 Ranking of tiger habitats by various centrality measures [34].**

Proceeding with the centrality analysis the most central patches included 3, 6, 7, 9, 13, 14, 17, 20, 21, 22, and 23 [34].

On correlation analysis using Pearson coefficients on all the centrality indices, the following correlation table were obtained:

|     | DC    | EC    | BC    | CC    | SC    | $SP_1$ | $SP_2$ | $SP_3$ | $SP_4$ | $SP_5$ | $SP_6$ | $SN_1$ | $SN_2$ | $SN_3$ | $SN_4$ | $SN_5$ | $SN_6$ |
|-----|-------|-------|-------|-------|-------|--------|--------|--------|--------|--------|--------|--------|--------|--------|--------|--------|--------|
| DC  | 1     |       |       |       |       |        |        |        |        |        |        |        |        |        |        |        |        |
| EC  | -0.16 | 1     |       |       |       |        |        |        |        |        |        |        |        |        |        |        |        |
| BC  | -0.28 | -0.1  | 1     |       |       |        |        |        |        |        |        |        |        |        |        |        |        |
| CC  | 0.044 | -0.24 | -0.01 | 1     |       |        |        |        |        |        |        |        |        |        |        |        |        |
| SC  | -0.05 | 0.493 | -0.23 | 0.054 | 1     |        |        |        |        |        |        |        |        |        |        |        |        |
| $SP_1$ | 0.046 | 0.521 | -0.3  | 0.122 | 0.96  | 1      |        |        |        |        |        |        |        |        |        |        |        |
| $SP_2$ | -0.08 | -0.11 | 0.182 | 0.449 | 0.112 | 0.054  | 1      |        |        |        |        |        |        |        |        |        |        |
| $SP_3$ | -0.21 | 0.278 | -0.03 | 0.297 | 0.199 | 0.258  | 0.335  | 1      |        |        |        |        |        |        |        |        |        |
| $SP_4$ | -0.23 | 0.264 | 0.253 | 0.151 | 0.238 | 0.218  | 0.314  | 0.767  | 1      |        |        |        |        |        |        |        |        |
| $SP_5$ | -0.11 | 0.239 | 0.212 | 0.205 | 0.175 | 0.216  | 0.273  | 0.784  | 0.942  | 1      |        |        |        |        |        |        |        |
| $SP_6$ | 0.206 | 0.257 | 0.263 | 0.237 | 0.032 | 0.081  | 0.324  | 0.589  | 0.674  | 0.726  | 1      |        |        |        |        |        |        |
| $SN_1$ | -0.02 | 0.546 | -0.14 | 0.028 | 0.92  | 0.934  | 0.058  | 0.252  | 0.292  | 0.273  | 0.129  | 1      |        |        |        |        |        |
| $SN_2$ | 0.106 | 0.742 | -0.17 | -0.06 | 0.824 | 0.846  | -0.01  | 0.16   | 0.223  | 0.197  | 0.221  | 0.904  | 1      |        |        |        |        |
| $SN_3$ | -0.18 | 0.635 | 0.067 | -0.16 | 0.341 | 0.327  | -0.05  | 0.157  | 0.188  | 0.143  | 0.299  | 0.421  | 0.497  | 1      |        |        |        |
| $SN_4$ | -0.22 | 0.641 | 0.095 | -0.2  | 0.355 | 0.307  | -0.01  | 0.127  | 0.206  | 0.129  | 0.288  | 0.408  | 0.492  | 0.99   | 1      |        |        |
| $SN_5$ | -0.19 | 0.927 | -0.04 | -0.27 | 0.59  | 0.576  | -0.05  | 0.161  | 0.306  | 0.247  | 0.22   | 0.643  | 0.794  | 0.688  | 0.714  | 1      |        |
| $SN_6$ | -0.16 | 0.975 | -0.15 | -0.28 | 0.562 | 0.59   | -0.04  | 0.292  | 0.279  | 0.254  | 0.227  | 0.616  | 0.767  | 0.66   | 0.666  | 0.952  | 1      |

**Fig. 4 Correlation table of Pearson coefficients for all centrality measures [34].**

A limitation in the modelling described in the paper [34] is that the corridor designing only captures optimality of a single graph with all PAs as the vertices and possible linkages between them as edges described over the entire landscape complex. A simplifying assumption in the work had been an absence of consideration of multiple possible corridors. The conclusive remarks mention, such a simplification is more often not in consonance with the real-world corridor scenario [33].

### 4. Modelling

The modelling in this study is both, a logical and a natural extension of the work explained in Section 3. The graph (algebraic representation of a network) $G_o = (V, E)$, represents the graph obtained in [34], where $V = \{1,2,3,...,23\}$; each integer in the set of vertices $V$ is a code allotted to the PAs in the Central India Landscape as shown in Table 1 and $E$ represents the set of edges between the ordered pair of vertices from set $V$. The graph $G_o$ is a particular instance of a simplex, and hence a network of vertices embedded in the new model

In the present study, we assume that the modelling presented in [34] for obtaining the minimum Cost path for designing the tiger corridor network in Central India Eastern Ghats landscape is valid, true and takes into account all the sufficient and important parameters to generate a weighted graph for obtaining a minimum spanning tree. The Protected Areas (PAs) in the landscape complex constitute the vertices and the collection of adjacencies within this complex that join any two of the PAs constitute the edges, comprising the focal landscape complex as a network. The existence of a passage between any two PAs represents some ecological flow, such as animal movement, between the adjacent PAS [34]. We thus assume that the Ecological flow of information and energy between any two connected PAs would be uniform on the network.

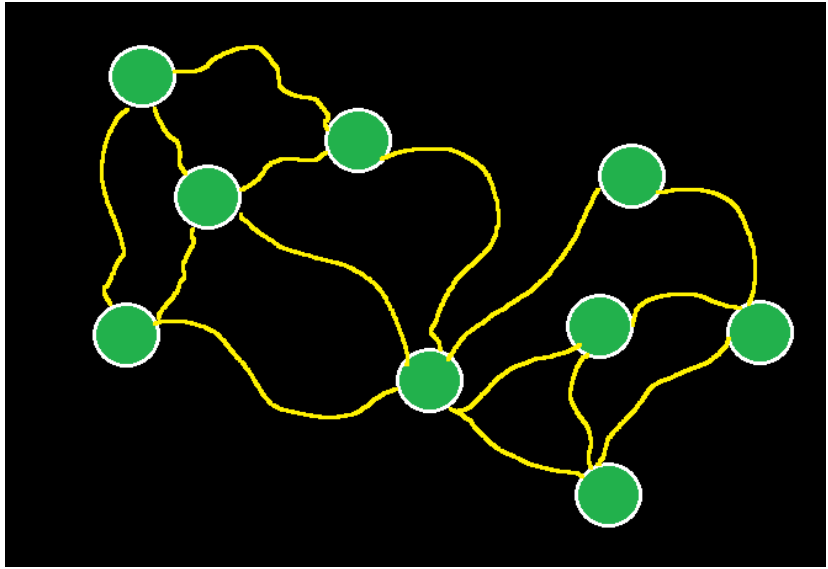

**Fig 5. Hypothetical landscape showing tiger PAs (solid green circular shapes), passage between the PAs (yellow curves joining the shapes) and the matrix (black colour background).**

In addition to the above discussion, we represent the landscape situation schematically using Fig. 5, where a rectangular frame represents the landscape, while the green circular shapes represent the PAs (habitats for tiger); with the connections between the habitats represented by the yellow curves. The black dark background in the figure represents the matrix, a component of the landscape that is neither patch nor corridor in the landscape [10]. The objective for the tiger is to compute an optimal as well as a complete path joining the different habitat patches, which would establish and promote a community structure satisfying its basic livelihood needs, minimizing the risk of its passage through the intervening landscape matrix [34].

The model corresponds to the map depicted in Fig. 8 of the Central India - Eastern Ghats landscape complex, spread over its constituent states [21]:

Wildlife Corridor in the landscape matrix are represented by graphs, where vertices correspond to the protected areas and edges to the possible structural and functional connections between them. The kind of graph is a weighted graph to illustrate the cost incurred by tigers (focal species) as a property for moving through the path. However, ordinary graphs do not adequately describe the migration structures through the networks. A key lacuna of the above-described theory is the lack of a convenient representation for the protected areas of delocalized movement patterns of tigers, which do not stay in the PAs and are territorial in nature.

With the above structural representation of corridors, if the problem is broken into sub-problems and a higher dimensional representation of the corridor network by simplicial complexes of the graphs, represent the structure. Simplicial complexes capture the essence on variations in the homomorphism of the landscape, which remains uncaptured while understanding the movement patterns of both tigers within protected areas as well as tigers, which are territorial in nature. For example, in the hypothetical landscape shown in Fig 5, a better approach to design the corridor networks would be to identify the community obtained from the centralities, forming a simplex to denote the structural as well as functional connectivity in the fragmented landscape providing higher degrees of freedom of movement to the focal species and a more complex strategy for conservation and mitigation of human animal conflicts in the area of interest. Fig 6 shows schemata for the design over hypothetical landscape. A higher dimensional proposed model captures more insights and factors, which improves the predictive ability of networks to identify the patterns and movement of tigers through the landscape. For the purpose of this modelling the work in this paper, we introduce simplexes as higher dimensional networks. Simplex structures enables one to discuss even multi-body interactions, instead of the necessarily two-body interactions that graphs allow for. Thus simplex structures provide computationally improved and mathematically robust approach to model the tiger corridor networks in the focal landscape.

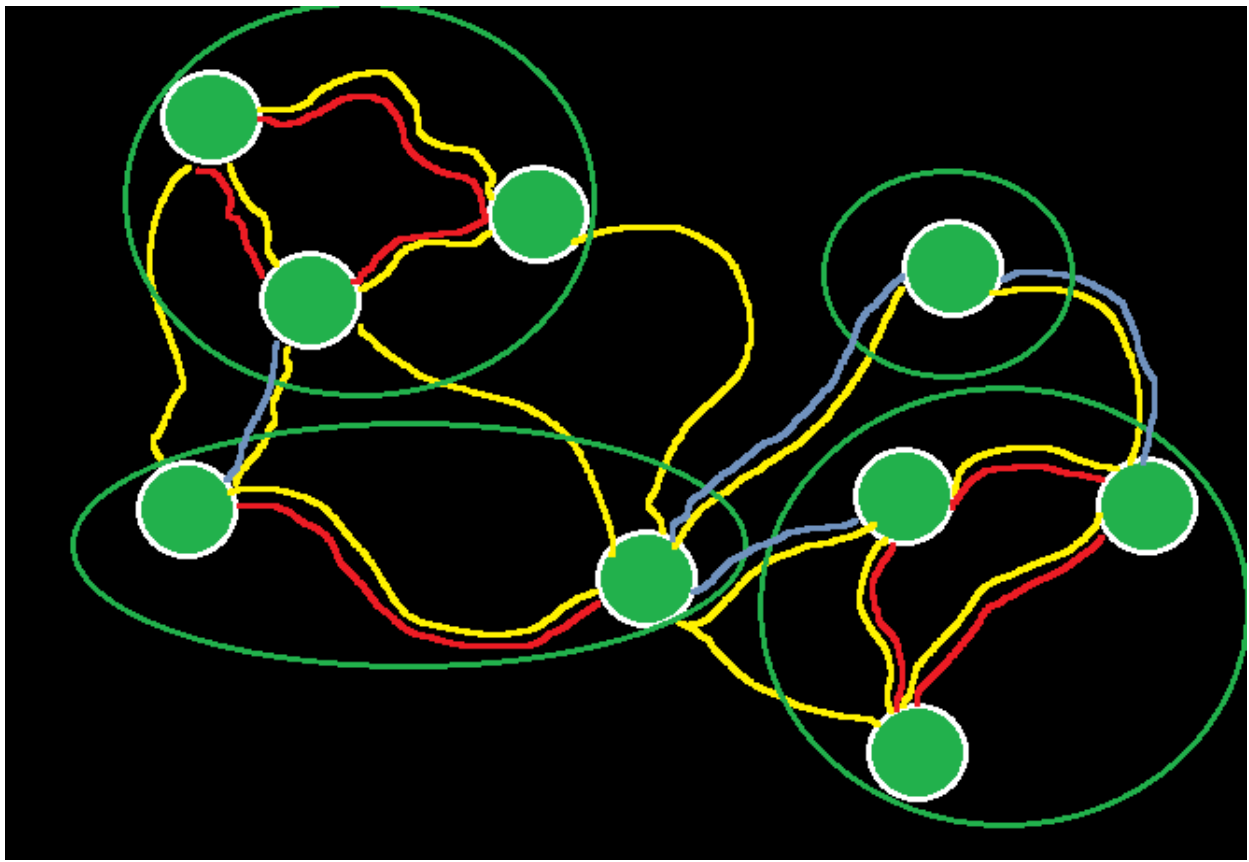

**Fig 6. Hypothetical landscape showing tiger PAs (solid green circular shapes), passage between the PAs (yellow curves joining the shapes), communities (green circular shapes), interconnection within the community (red curves joining the solid green shapes), interconnection between different communities (grey curves joining the solid green shapes) and the matrix (black color background).**

With reference to the above discussion, using the Newman – Girvan algorithm, the following communities in the network, have been identified and generalized (The elements in the set correspond to the PAs, encoded as shown in Table 1) [34]:

Community 1 :{ 1, 2, 3, 4, 22}

Community 2 :{ 11}

Community 3 :{ 5, 6, 7}

Community 4 :{ 8, 9, 10, 12, 13, 21, 23}

Community 5 :{ 14, 17, 18, 19, 20}

Community 6 :{ 15, 16}

Based on the above community identifications done through centrality measures and corresponding analysis, obtained from the modellings done in the previous work, in the present work we identify community that would form a simplex [50], denoted below in the Table with the simplex structure and the vertices belonging to those simplexes:

**Table 2: Simplex Structure for the Tiger Corridor Networks in Central India Eastern Ghats Landscape**

| Community | k-Simplex | Vertices |
|---|---|---|
| | | |
| 1 | 4-Simplex | {1,2,3,4,22} |
| 2 | 0-Simplex | {11} |
| 3 | 2-Simplex | {5,6,7} |
| 4 | 6-Simplex | {8,9,10,12,13,21,23} |
| 5 | 4-Simplex | {14,17,18,19,20} |
| 6 | 1-Simplex | {15,16} |

Movement of tigers in the designed corridors is independent of directions and dependent on the availability of resources like prey-base, water, etc. [36]. Hence, in order to construct a more realistic model of the tiger corridors in the present landscape, we create a simplicial complex, which would present a model ensuring efficiency with tigers of PAs and the territorial tigers using higher degrees in the focal landscape by considering all possible linkages. For the mentioned purpose, we compute the upper adjacencies for each k-simplex based on the (k-1) faces belonging to the simplex, which denote the generalization of movement of the species within the above-defined community presented in Table 3. The Upper adjacencies help to design the Tiger corridor network within the same community providing more pathways for movement with higher functional features but less constrained on the structural parameters [47].

**Table 3: Upper Adjacencies for the Tiger Corridor Simplex in Central India Eastern Ghats Landscape**

| Community | Vertices/Vertex | Upper Adjacency Sets |
|---|---|---|
| 1 | {1,2,3,4,22} | {1,2,3,4}, {1,2,3,22}, {1,3,4,22}, {1,2,4,22}, {2,3,4,22} |
| 2 | {11} | { } |
| 3 | {5,6,7} | {5,6}, {6,7}, {5,7} |
| 4 | {8,9,10,12,13,21,23} | {8,9,10,12,13,21}, {8,9,10,12,13,23}, {8,9,10,12,23,21}, {8,9,10,23,13,21}, {8,9,23,12,13,21}, {8,23,10,12,13,21}, {{23,9,10,12,13,21} |
| 5 | {14,17,18,19,20} | {14,17,18,19}, {14,17,18,20}, {14,17,19,20}, {14,18,19,20}, {17,18,19,20} |
| 6 | {15,16} | {15}, {16} |

The Lower adjacencies between the major simplex containing all the 23 vertices is calculated in order to obtain the structural connectivity between all the possible communities calculated through the Upper adjacencies, based on the connections between various communities and the key vertices which play as the connecting vertices based on Betweenness centrality measures and rankings, presented in Table 4.

**Table 4: Lower Adjacencies for the Tiger Corridor Simplex in Central India Eastern Ghats Landscape**

| Lower Adjacency Sets |
|---|
| {16} ⌣ {18} |
| {22} ⌣ {20} |
| {4} ⌣ {5} |
| {4} ⌣ {6} |
| {7} ⌣ {23} |
| {12} ⌣ {11} |
| {21} ⌣ {23} |
| {21} ⌣ {13} |
| {21} ⌣ {20} |
| {21} ⌣ {14} |
| {21} ⌣ {17} |

## 5. Result and Discussion

A strong argument and theory established in this paper suggests that migrating species during movement from one habitat patch to the other may not always take up the most optimal path, which can be proposed and found out on the application of existing algorithms. Individuals at different times may have different specific requirements and may take a path that instantaneously satisfies those demands [43]. Hence, consideration of optimal paths for designing of wildlife corridors must be a basic computational framework but not always represent the only paths, which species would consider for movement.

The modelling addresses the adjacency of different communities by extracting a graph treating each community as a vertex and all possible paths between them as the edges. The vertices in the graph $G_o$ create six communities as shown in Table 2. A new graph $G_c = (V_c, E_c)$ created from $G_o$, where $V_c = \{C_1, C_2, C_3, C_4, C_5, C_6\}$; $C_i$: community it obtained over the graph $G_o$ $\forall\ i = \{1,6\}$. On application of Lower adjacencies, linkages between all the vertices of $G_c$ implies that the most optimal edges between the $C_i$ have been obtained. On application of upper adjacency, the proposed solution obtains the adjacency of different PAs by extracting a graph treating each PA as a vertex and all possible paths between them as the edges. Six non-isomorphic graphs $G_1$, $G_2$, $G_3$, $G_4$, $G_5$, $G_6$, are obtained where each graph $G_i$ represents the community $i$. The set of vertices of each graph represent the PAs and the set of edges represent the territorial regions, which would facilitate the movement of tigers.

For the problem in the paper, the elements, that provide solutions, include inter-community and intra-community movements. The former suggests that within a community, there exist many possible paths, which animals may use for shorter geographical movements and in search of possible paths, which could take them to a suitable habitat patch for survival and other ecological activities [44]. Following from the Theorem 1, mentioned in Section 2 of the paper, the theory of a landscape linkage in each community satisfies the mapping of every possible connection within the communities and hence is complete. Whereas the later suggests that when the greater structural components of corridors are considered, animals may not deviate migrating for larger distances from a path, which is not both structurally and functionally optimal and connected [45].

To present the above discussions, we have used the concept of simplicial complexes to model the community structures. Upper adjacencies model the intra-community structures and lower adjacencies connect the communities with each other and model the inter-community structures [46].

In the proposed model, within a community, each PA acts as a vertex and various degrees of connectivity for each vertex represents the different faces of the clique complex [49]. Each face presents a path, which Tigers may consider for migration from the source-protected areas or the source vertex. As each face represents a separate simplicial complex until we receive 0-complex, which would be a single path for the tigers hence within a complex, tigers get n numbers of path combination for movement, where n represents the degree of the community simplex.

For the inter-community movement there are vertices i.e. the PAs within the communities that connect to the vertices or Protected Areas from a different community and hence provide movement of Tigers from one community to the other. Lower adjacencies in the simplex provide these connections of vertices between different communities.

The simplex that derives the connection of various sub-graphs for the Tiger Corridor Network in Central India Eastern Ghats Landscape is shown in Fig 7.

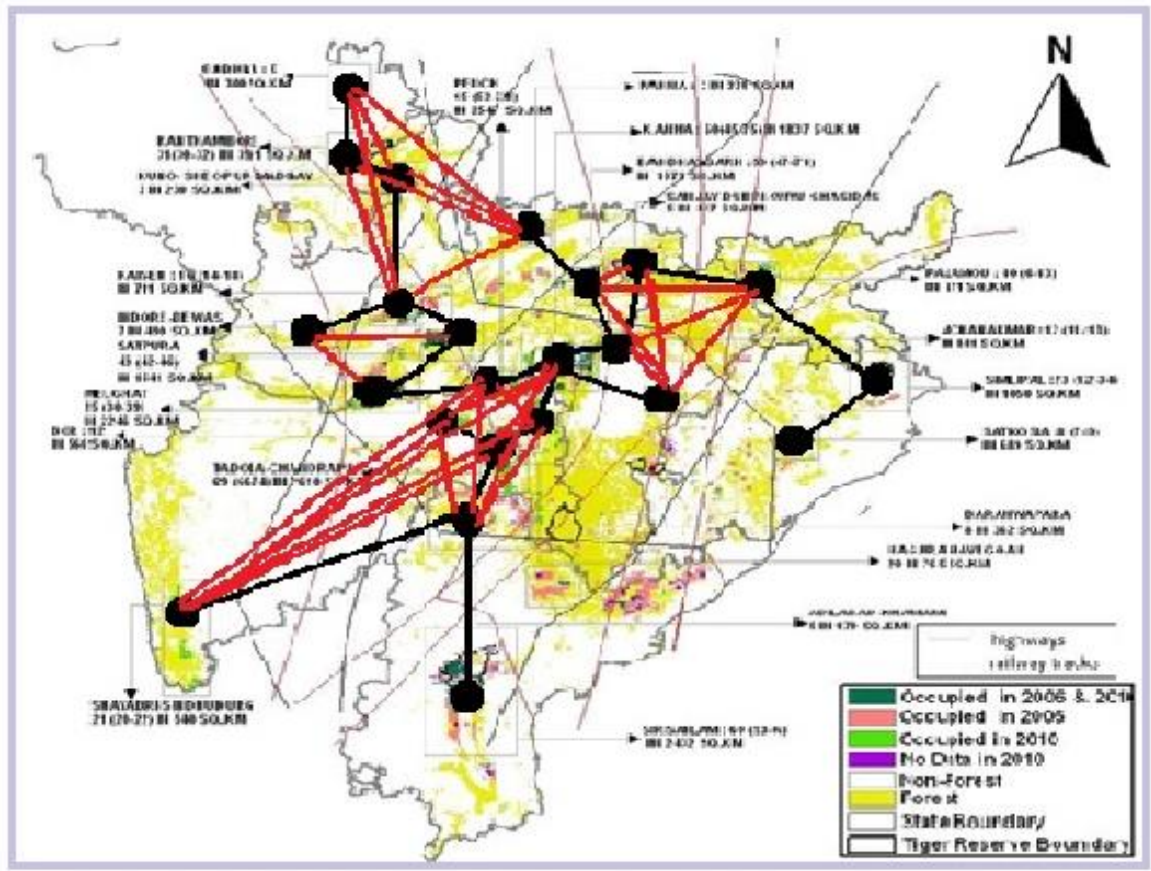

(a)

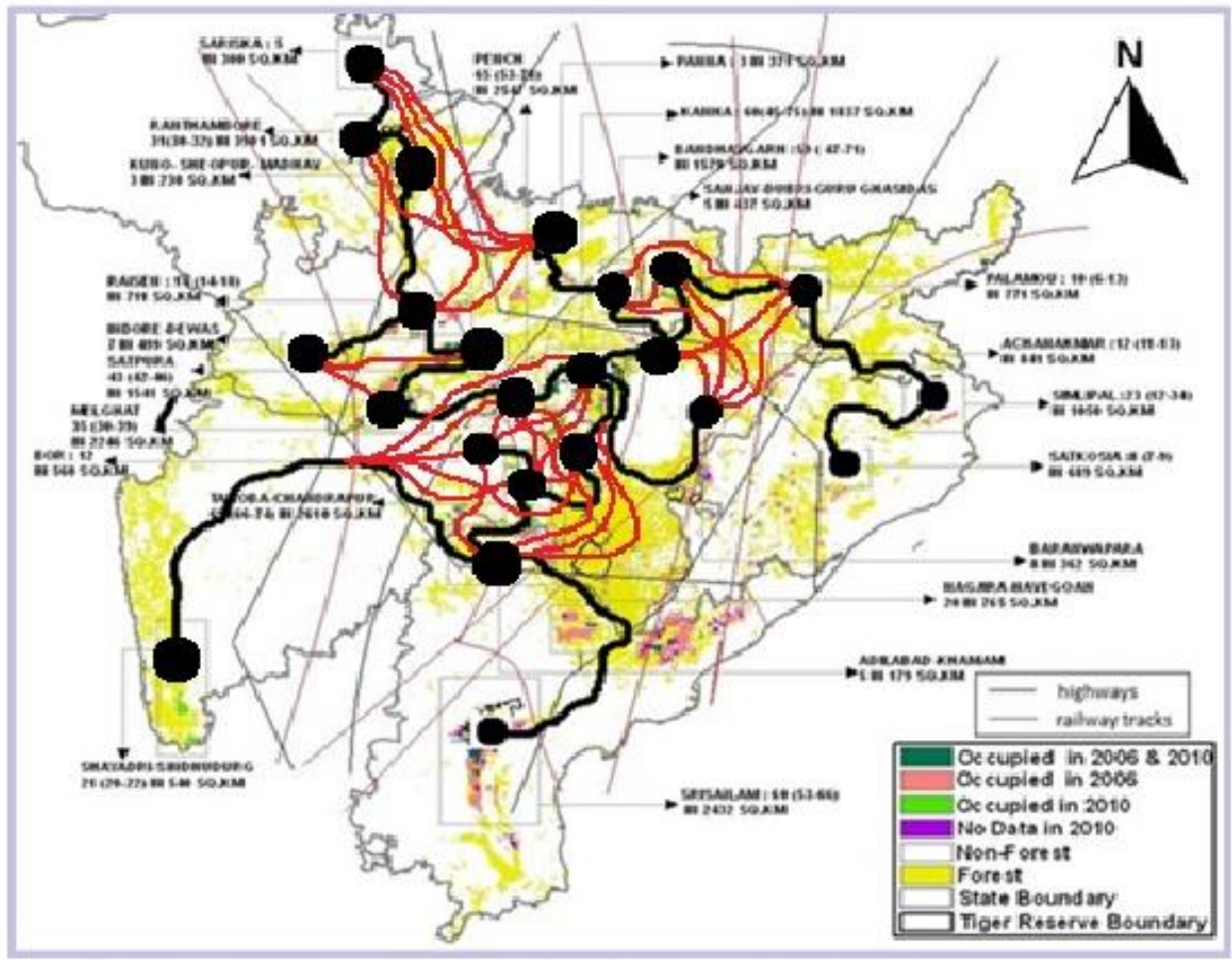

(b)

**Fig. 7 A feasible tiger corridor network, given by a Simplicial Complex obtained over the MST presented in [34], (a) represents straight-line cliques for vertex linkages in the landscape and (b) overlaid on the map of the focal landscape complex.**

The result obtained through the modelling and presented in Fig. 7 have been compared with the results presented in [34] and the field observations obtained from literatures and recent human-animal conflict activities. The network simplex algorithm moves from one feasible spanning tree structure to another until it obtains a spanning tree structure that satisfies the optimality condition [19]. Hence, comparative studies are presented in Table 5 and Figure 8 as shown:

Table 5: The highlighted in yellow represent proposed corridors in [34]; the cells highlighted in green represent the new addition of corridors [48] from the present model to the proposed corridors of [34] and the Blue Highlighted patches represent the Field observed human-animal conflict regions.

| Protected Areas(Habitat Patch) | 1 | 2 | 3 | 4 | 5 | 6 | 7 | 8 | 9 | 10 | 11 | 12 | 13 | 14 | 15 | 16 | 17 | 18 | 19 | 20 | 21 | 22 | 23 |
|---|---|---|---|---|---|---|---|---|---|---|---|---|---|---|---|---|---|---|---|---|---|---|---|
| 1 |   | 1 | 1 | 1 |   |   |   |   |   |   |   |   |   |   |   |   |   |   |   |   |   | 1 |   |
| 2 | 1 |   | 1 | 1 |   |   |   |   |   |   |   |   |   |   |   |   |   |   |   |   |   | 1 |   |
| 3 | 1 | 1 |   | 1 |   |   |   |   |   |   |   |   |   |   |   |   |   |   |   |   |   | 1 |   |
| 4 | 1 | 1 | 1 |   | 1 | 1 |   |   |   |   |   |   |   |   |   |   |   |   |   |   |   | 1 |   |
| 5 |   |   |   | 1 |   | 1 | 1 |   |   |   |   |   |   |   |   |   |   |   |   |   |   |   |   |
| 6 |   |   |   | 1 | 1 |   | 1 |   |   |   |   |   |   |   |   |   |   |   |   |   |   |   |   |
| 7 |   |   |   |   | 1 | 1 |   |   |   |   |   |   |   |   |   |   |   |   |   |   |   |   | 1 |
| 8 |   |   |   |   |   |   |   |   | 1 | 1 |   | 1 | 1 |   |   |   |   |   |   |   | 1 |   | 1 |
| 9 |   |   |   |   |   |   |   | 1 |   | 1 |   | 1 | 1 |   |   |   |   |   |   |   | 1 |   | 1 |
| 10 |   |   |   |   |   |   |   | 1 | 1 |   |   | 1 | 1 |   |   |   |   |   |   |   | 1 |   | 1 |
| 11 |   |   |   |   |   |   |   |   |   |   |   | 1 |   |   |   |   |   |   |   |   |   |   |   |
| 12 |   |   |   |   |   |   |   | 1 | 1 | 1 | 1 |   | 1 |   |   |   |   |   |   |   | 1 |   | 1 |
| 13 |   |   |   |   |   |   |   | 1 | 1 | 1 |   | 1 |   |   |   |   |   |   |   |   | 1 |   | 1 |
| 14 |   |   |   |   |   |   |   |   |   |   |   |   |   |   |   |   | 1 | 1 | 1 | 1 | 1 |   |   |
| 15 |   |   |   |   |   |   |   |   |   |   |   |   |   |   |   | 1 |   |   |   |   |   |   |   |
| 16 |   |   |   |   |   |   |   |   |   |   |   |   |   |   | 1 |   |   | 1 |   |   |   |   |   |
| 17 |   |   |   |   |   |   |   |   |   |   |   |   |   | 1 |   |   |   | 1 | 1 | 1 | 1 |   |   |
| 18 |   |   |   |   |   |   |   |   |   |   |   |   |   | 1 |   | 1 | 1 |   | 1 | 1 |   |   |   |
| 19 |   |   |   |   |   |   |   |   |   |   |   |   |   | 1 |   | 1 | 1 | 1 |   | 1 |   |   |   |
| 20 |   |   |   |   |   |   |   |   |   |   |   |   |   | 1 |   |   | 1 | 1 | 1 |   | 1 | 1 |   |
| 21 |   |   |   |   |   |   |   | 1 | 1 | 1 |   | 1 | 1 | 1 |   |   | 1 |   |   | 1 |   |   | 1 |
| 22 | 1 | 1 | 1 | 1 |   |   |   |   |   |   |   |   |   |   |   |   |   |   |   |   | 1 |   |   |
| 23 |   |   |   |   |   |   | 1 | 1 | 1 | 1 |   | 1 | 1 |   |   |   |   |   |   |   | 1 |   |   |

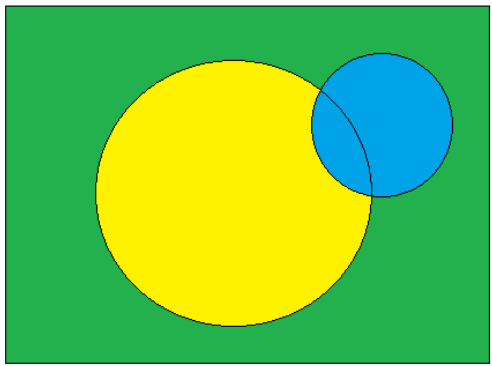

**Fig. 8.** The green region represents the proposed corridors through the simplicial complex model, which also includes the proposed paths through the the minimum spanning tree model represented by the yellow sphere and the blue sphere represents the animal-human conflict regions found from the literature survey, expert reports and news articles which shares some common region with the proposed corridors in the minimum spanning tree concept and some corridors which were not identified by the model but have been identified by the simplicial complex model.

## 6. Conclusion

The present work has been developed with objectives to (i) obtain a feasible tiger corridor network which could capture the movement patterns of tiger present in both, the PAs as well as the territorial population of the landscape, (ii) to identify the most important habitat patches, along with their underlying community structure. These objectives were decided in order to devise a formal framework that would address the shortcomings of [34].

The limitation in the modelling described in [34] is that the corridor designing addressed the structural definition of connectivity and focused only on the PAs containing tiger population; thereby ignoring the territorial populations of the species. In the present work, the results obtained through simplicial complex modelling in ecological landscape capture both structural and functional features of the focal landscape for designing the corridors. Further, improving on the basic computational framework provided in [34], the present modelling uses centralities to obtain community and then have modelled and realized the corridor network as simplicial complexes. A simplifying assumption in the previous work had been an absence of consideration of multiple possible paths between connected PAs. In the present work, multiple paths between connected PAs, considered as faces of the simplex, improved on designing more facet approach of the computational model for perceiving a viable corridor network. Finally, in the previous work, the authors justified this absence of path redundancy consideration due to two reasons: first, priority in the paper was to focus on network efficiency over redundancy, and second, the work focused on estimation of optimal spanning tree connecting the PAs, rather than inclusion of alternative paths [33, 34]. With a changed priority and a primary objective to capture the interactions between the members of both kinds of tiger population: the PA tiger population and the territorial tiger population with identification of the most important habitat patches, along with their underlying community structure, we present a more realistic model including redundant connectivity with inclusion of alternate paths.

Thus, the work presented in this paper presents an improved approach of previous works presented in [30, 34]. The potential corridors identified by the proposed model in the focal landscape complex using the concept of simplicial complexes are not all least resistance paths.

The illustrations through a Venn diagram in Fig 8, shows that the approach of designing wildlife corridor networks is more robust with the mathematical theory of simplicial complexes. The work also indicates the existence of territorial patches as internal vertices in the tiger corridors, as implied by the edge-betweeness centrality. These patches, though may not have the capacity to hold permanent tiger populations yet can act as temporary shelter and refuge for transiting tigers particularly during the daytime.